\documentclass[5p]{elsarticle}

\usepackage[utf8]{inputenc}
\usepackage{caption}
\usepackage{graphicx}
\usepackage{bm}
\usepackage{float}
\usepackage{makeidx}
\usepackage{ifpdf}
\usepackage{url}
\usepackage{amssymb}
\usepackage{color}
\usepackage{amsmath}

\usepackage{soul}

\usepackage{lineno,hyperref}
\modulolinenumbers[10]

\begin{document}
\sloppy

\begin{frontmatter}

\title{On identifying dynamic length scales in crystal plasticity}

\author[ELTEaddress]{D\'{e}nes Berta}
\author[TAUaddress]{David Kurunczi-Papp}
\author[TAUaddress]{Lasse Laurson\corref{mycorr}}\ead{lasse.laurson@tuni.fi}
\author[ELTEaddress]{P\'{e}ter Dus\'an Isp\'{a}novity\corref{mycorr}}\ead{ispanovity.peter@ttk.elte.hu}
\cortext[mycorr]{Corresponding authors}

\address[ELTEaddress]{Department of Materials Physics, ELTE E\"{o}tv\"{o}s Lor\'{a}nd University, P\'{a}zm\'{a}ny P\'{e}ter s\'{e}tany 1/a, 1117 Budapest, Hungary}

\address[TAUaddress]{Computational Physics Laboratory, Tampere University, P.O. Box 692, FI-33014 Tampere, Finland}

\date{\today}

\begin{abstract}
Materials are often heterogeneous at various length scales, with variations in grain structure, defects, and composition which has a strong influence on the emergent macroscopic plastic behavior. In particular, heterogeneities lead to fluctuations in the plastic response in the form of jerky flow and ubiquitous strain bursts. One of the crucial aspects of plasticity modeling is scale bridging: In order to deliver physically correct crystal plasticity models, one needs to determine relevant microstructural length scales. In this paper we advance the idea that continuum descriptions of dislocation mediated plasticity cannot neglect dynamic correlations related to the avalanche behavior. We present an extensive weakest link analysis of crystal plasticity by means of three-dimensional discrete dislocation dynamics simulations with and without spherical precipitates. We investigate strain bursts and related length scales and conclude that while sufficiently strong obstacles to dislocation motion tend to confine the dislocation avalanches within well-defined sub-volumes, in pure dislocation systems the avalanches may span the system, implying that the dynamic length scale is, in fact, the size of the entire sample. Consequences of this finding on continuum modeling are thoroughly discussed.
\end{abstract}


\end{frontmatter}


\section{Introduction}

During plastic deformation of crystalline matter dislocations move and interact in a complex manner due to their anisotropic long-range stress fields, constrained motion and various short-range interactions. Modeling and understanding this complex spatio-temporal dynamics is vital for predicting the mechanical behavior of materials. Representing the dislocation microstructure can be performed on multiple scales: i) molecular dynamics (MD) considers every atom in the crystal, ii) discrete dislocation dynamics (DDD) considers every discrete dislocation in the system as line-like objects and represents the atomic lattice with an elastic continuum, while iii) continuum dislocation dynamics (CDD) models homogenize the dislocations as well and are formulated on the level of continuous dislocation density fields \cite{bulatov2006computer,bertin2020frontiers}. Whereas the main issue in the MD $\to$ DDD transition are the mobility rules of individual dislocations \cite{PhysRevE.93.013309,sobie2017scale,bertin2023learning}, the fundamental quantities in the DDD $\to$ CDD scale bridging are length scales related to the microstructure \cite{groma2003spatial,hochrainer2022making}. One of the reasons for the latter is that spatial correlations build up between dislocations (one form of which are dislocation patterns) that one can take into account in CDD theories with strain gradients and corresponding length-scales related to \emph{static} correlation lengths \cite{hochrainer2016thermodynamically,anderson2024dislocation}. The term ``static'' corresponds to the fact, that these length parameters are obtained from instantaneous snapshots of the dislocation microstructure \cite{zaiser2001statistical, sills2018dislocation, hochrainer2022making, anderson2024dislocation}.

Since this paper is concerned with \emph{dynamic} length scales that are related not to individual states of a dislocation structure, rather its dynamics, let us now summarize some important features related to dislocation motion. Recently it was found that crystal plasticity is governed by wildly fluctuating mechanisms characterized by broad, power-law-like distributions of acoustic emission events that correspond to the sudden rearrangement of dislocations~\cite{weiss2003three,zaiser2006,alava2014crackling,weiss2015mild,papanikolaou2017avalanches}. Deformation, thus, proceeds in a sequence of short plastic events, characterized by a sudden unzipping of dislocation sub-structures in a certain sub-volume of the crystal followed by their quick rearrangement~\cite{csikor2007dislocation,szabo2015plastic,Sparks2018}. When the sample size is small (typically few microns or below) these events, called dislocation avalanches, may span the entire specimen and, therefore, manifest in large and unpredictable strain bursts that prevent controlled forming. This phenomenon, therefore, poses a significant challenge for practical applications at this size scale.

The main strategy to mitigate the negative effects of these bursts is to control their maximum size. As it was shown experimentally by Weiss and co-workers~\cite{weiss.2021r}, introducing quenched disorder in the crystal lattice (through, for instance, precipitation or alloying) may limit the spatial extension of the fluctuations by obstructing the motion of the dislocations~\cite{ardell,papanikolaou2017obstacles,beyerlein2019,ovaska2015quenched,laurson2024dislocation}. It was suggested, that the phenomenon is related to a (dynamic) length scale: whenever the sample is smaller than this scale plastic fluctuations are \emph{wild} (that is, avalanches span the whole sample and forming cannot be controlled due to large strain bursts), and if the sample is larger than this scale, then fluctuations become \emph{mild} (that is, the plastic response is bulk-like). Nevertheless, the physical background of this process and how such a dynamic length scale could be introduced are still elusive. We stress, however, that in both regimes dislocations move in a stick-slip manner, that is, the avalanche-like motion of dislocations is ubiquitous, the difference lies mainly in the avalanche sizes and their distribution.

The main objective of the research presented in this paper is to investigate the spatial extension of dislocation avalanches and, in particular, the role of precipitates in confining the avalanches spatially. More specifically, we focus on characterization of the extent of dynamic correlations at the onset of the first dislocation avalanche, i.e., the ``excitation volume". This issue is naturally connected to the question of whether plasticity is a weakest link phenomenon~\cite{parthasarathy2007contribution,lai2008bulk,ELAWADY200932,ye2010extraction,ISPANOVITY20136234,derlet2015,derlet2016,ispanovity2017role,gelebart2021grain}, as well as to the nature of the excitation spectrum quantified by presence or absence of singularities in the distribution of local excitation stresses~\cite{ovaska2017excitation}. It is a natural assumption that if a solid body is subjected to external stress, plastic events (dislocation avalanches) will be triggered at the subsequent weakest spots of the material. This idea was also supported by experimental observations, like weakest link statistics of micropillar strengths, motivating several successful models of plasticity in the recent years~\cite{zaiser2005fluctuation,papanikolaou2012quasi}. This approach suggests that dislocation avalanches can be triggered individually and are characterized by a local volume being excited at a certain yield stress. This means that whenever mechanical load is applied to a part of the sample larger than the excitation volume, the avalanche can be triggered. For triggering the applied stress must exceed a given threshold, that can be considered as the yield stress of that sub-volume of the material. So it is straightforward to assume, that microstructural inhomogeneities of materials can be represented in a higher-scale description by \emph{local yield stresses} (LYSs). Although this assumption seems straightforward, in a simplified 2D discrete dislocation dynamics (DDD) model it turned out that the conventional weakest link argumentation is only valid in systems with obstacles for dislocation motion~\cite{berta2023dynamic}. In pure (obstacle-free) systems, avalanches could always span the entire volume due to the $1/r$-type long-range interaction of dislocations~\cite{ispanovity2014avalanches}. In conclusion, in the terminology of wild and mild fluctuations mentioned above, pure systems are always wild and the obstacles with short-range stress fields introduce a length scale and may make fluctuations mild \cite{ovaska2015quenched}.

However, 2D DDD models with point-like (that is, always straight) dislocations cannot, by nature, properly describe all the features of realistic 3D dislocation dynamics. This applies especially to the description of dislocation lines interacting with point-like obstacles (e.g., spherical precipitates), including phenomena such as bowing out of dislocation lines around and in between precipitates as well as formation of Orowan loops, which obviously require a full 3D model. Hence, here we aim at addressing the pertinent open question of whether and how the above considerations on the local strengths and the weakest link behavior extend to the realistic case of 3D DDD simulations with and without precipitates. In particular, we intend to determine the LYS distributions in 3D DDD simulations and address how a dynamic length scale can be introduced and how it depends on the presence of static point-like obstacles.

The paper is organized as follows: In the following section (Section \ref{sec:methods}), we present the methodology used, i.e., the 3D DDD simulations with local loading, describing also the procedure of detecting the first avalanche. Section~\ref{sec:results} then presents the results of our study, focusing in particular on the scale-dependent distributions of the LYS as well as on an analysis assessing to what extent these obey the weakest link picture. The paper is finished in Section~\ref{sec:conclusions} with discussion and conclusions.

\section{Methodology}
\label{sec:methods}
\subsection{DDD simulations}
The simulations presented in this paper are performed using a modified version of the 3D DDD software ParaDiS~\cite{arsenlis2007enabling}, with the additional inclusion of spherical obstacles or precipitates~\cite{PhysRevE.93.013309}. In ParaDiS the dislocation lines are discretized into nodal points connected by straight line segments. Dislocation motion is realized by moving the nodal points, which, additionally, can be added or removed depending on segment lengths and curvatures. The total Peach-Koehler force acting on a node is calculated from the external stress, the internal stress field generated by the dislocation segments within the crystal and dislocation line tension~\cite{Dislocations}. The forces generated by the internal fields are divided into local and far-field ones. The local forces are computed directly via line integrals, while the far-field forces are obtained from the course-grained dislocation structure using a fast multipole method (FMM)~\cite{greengard_rokhlin_1997}. FMM works without the need of a cut-off distance and is widely used in other physical systems that contain long-range forces, such as Coulomb interactions in ionic crystals or gravitational fields in galaxies. The calculation of the forces is followed by the time integration and solving the equations of motion for the discretized nodes. ParaDiS takes into account the crystal structures and dislocation character (edge, screw, or a combination of these). The material specific properties enter the equations of motion through the mobility functions, relating the forces with the dislocation velocities $\bm v = M \bm F_\mathrm{glide}$, where $M$ is the dislocation mobility and $\bm F_\mathrm{glide}$ is the glide component of the Peach-Koehler force. Bulk properties are simulated by using periodic boundary conditions (PBC). The interactions of the dislocation segments in the simulation box with all of their periodic images is handled within the FMM algorithm.

\begin{figure*}[ht!]
    \centering
    \includegraphics[width=\textwidth]{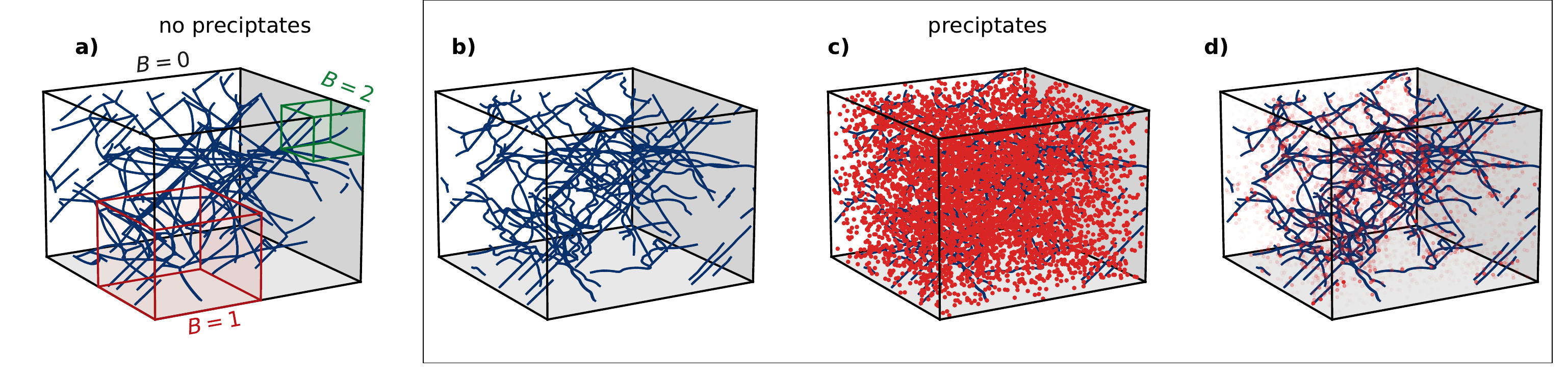}
    \caption{(a) A dislocation configuration without precipitates stressed to $100\,\mathrm{MPa}$ with examples of subboxes for $B=1$ and $B=2$. (b) A configuration including precipitates (which are not shown) stressed to the same value. Note the increased roughness of dislocation lines caused by the presence of precipitates. (c) The same configuration with the precipitates also shown. (d) The same configuration with precipitate opacity proportional to the distance from the closest dislocation line. This plot emphasizes the role of interaction with precipitates in the roughening of the dislocation structure.}
    \label{fig:congfigs2}
\end{figure*}

The modification of ParaDiS incorporates a description of immobile spherical obstacles or precipitates. The dislocation-precipitate interaction is described by the radially symmetric Gaussian potential
\begin{equation}\label{eq:precipitate}
    \bm F(\bm r)=-\nabla U(\bm r)=-\nabla\frac{2Ab^3r\mathrm{e}^{-\frac{r^2}{r_\mathrm p^2}}}{r_\mathrm p^2}
\end{equation}
with the Burgers vector $b$, the interaction strength $A$ and the precipitate radius $r_\mathrm{p}$. In general, the Gaussian potential allows tuning of the defect strength from weak, shearable obstacles ($A\lesssim10^{10}\,\mathrm{Pa}$) to strong impenetrable obstacles ($A\gtrsim10^{11}\,\mathrm{Pa}$), which the dislocations have to bypass via the Orowan mechanism~\cite{PhysRevE.93.013309,salmenjoki2020plastic}.

We consider here an elastically isotropic single crystalline material with FCC structure and parameters of Al: shear modulus $G=26\,\mathrm{GPa}$, Poisson ratio $\nu=0.35$, Burgers vector $b=2.863\times10^{-10}\,\mathrm{m}$ and, for simplicity, the same dislocation mobility $M=10^4\ \mathrm{Pa}^{-1}\mathrm s^{-1}$ in all directions. The linear system size $L=1.43\,\mathrm{\mu m}$ (typical for microcrystal compression experiments~\cite{dimiduk2006scale}) is chosen in a way that $N_0=40$ initially straight dislocation lines placed randomly on the glide planes of the FCC lattice relax at zero applied stress to a dislocation configuration in a metastable state with the approximate density $\rho_0\approx2.5\times10^{13}\,\mathrm{m}^{-2}$. The systems with strong obstacles ($A=10^{11}\,\mathrm{Pa}$) were created by randomly placing $N_\mathrm p=5000$ precipitates with radius $r_\mathrm p=40b$ (corresponding to $\sim 1\%$ volume fraction typical of aluminium alloying) into a previously relaxed pure dislocation configuration. Here a second relaxation step was performed so the dislocation configuration could respond to the presence of the newly introduced precipitates. After the relaxation stages, a quasistatic stress-controlled loading was implemented, where the stress is either increased with the constant rate $\dot{\sigma}=10^{15}\,\mathrm{Pa}/\mathrm{s}$ as long as the collective mean dislocation velocity is below a threshold or is kept constant with accumulating plastic strain when the velocity is above the threshold. Notice that in the system with precipitates, the mean precipitate spacing sets a length scale that breaks the similitude principle of pure dislocation systems~\cite{zaiser2014scaling}.

\begin{figure*}[ht!]
    \centering
    \includegraphics[width=\textwidth]{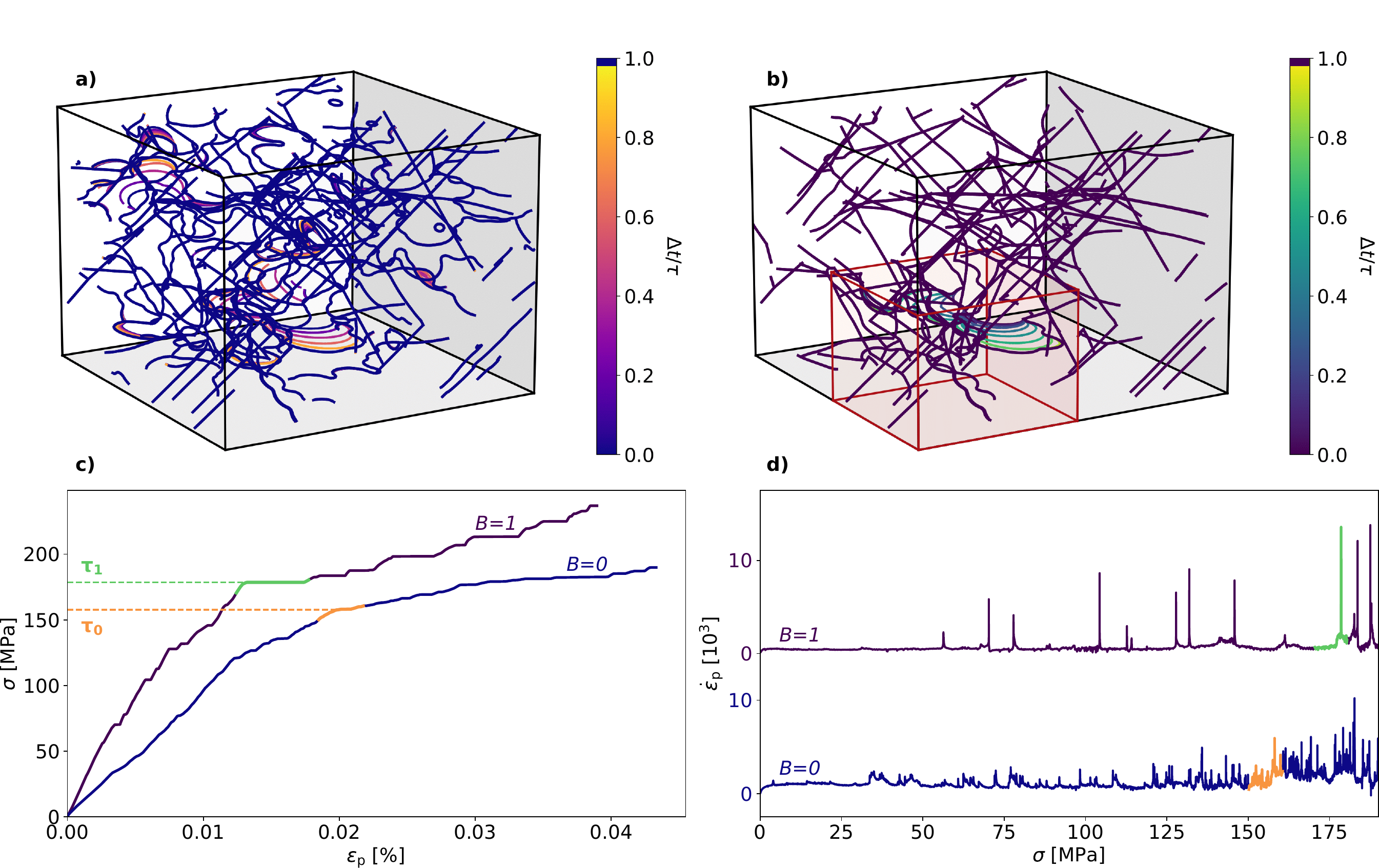}
    \caption{(a) The evolution of the dislocation structure in a configuration with precipitates (not shown) during a dislocation avalanche. In this case the whole simulation box is loaded ($B=0$). The color scale $\Delta t/\tau$ is the dimensionless time which is 0 at the beginning and 1 at the end of the plastic event. (b) The evolution of the dislocation structure of the same configuration when only a small cubic region (highlighted in red) is loaded ($B=1$). The dislocation motion within this region is very similar to the local motion in the $B=0$ case. (c) The stress-plastic strain curves corresponding to (a) and (b) with a color code consistent with the 3D plots. The two events that are alike in appearance occur at a similar external stress (within $15\%$ difference) as well. (d) The plastic strain rate $\dot{\varepsilon}_\mathrm{p}$ as a function of loading stress $\sigma$ for the systems corresponding to (a) and (b). Due to the enhanced dislocation motion the avalanches manifest in peaks in the plastic strain rate. The different noise level and dissimilar peak values of the plastic strain rate make the use of different thresholds for different $B$ values sensible. }
    \label{fig:3d}
\end{figure*}

\subsection{Local loading}

In order to characterize the local strength of sub-volumes, we utilize the concept of LYS, that is, the stress threshold of yielding during loading a small local sub-volume. This local variable was shown to correlate very strongly with the loci of plastic events during global loading in model amorphous solids \cite{patinet2016connecting, barbot2018local,ruan2022predicting}. The LYS also proved to be an important variable in mesoscale simulations~\cite{baret2002extremal, talamali2011avalanches, liu2016driving, budrikis2017universal}, in continuum dislocation field theories ~\cite{ispanovity2017role, ispanovity2020emergence, wu2021cell}, and in nanoindentation experiments used for measuring local hardness as well~\cite{SHEN2016171,song2019discrete}. Our loading protocol used to measure the LYS is motivated by the one used to probe model amorphous materials \cite{patinet2016connecting, barbot2018local,  richard2020predicting} and is similar to the method applied earlier in 2D discrete dislocation systems \cite{ovaska2017excitation, berta2023dynamic, berta2024avalanche}. The LYSs are measured in cubic (sub)systems (see Fig.~\ref{fig:congfigs2}(a) and Fig.~\ref{fig:3d}(b)) to which we will refer to as boxes. These boxes are obtained by a recursive division of the system. The number of subsequent division steps is denoted by $B$. $B=0$ corresponds to the whole simulation cell, $B=1$ consists of its $2^3=8$ boxes after one division and $B=2$ corresponds to the $8\times 8=64$ boxes after two subsequent steps of division. During local probing the quasi-static loading protocol described in the previous subsection is used in every case. If $B>0$, local loading is realized through dislocation mobility: dislocation nodes outside the box are immobile ($M=0$) while the ones within the box are let to move normally.

\subsection{Avalanche detection}

The avalanche detection is based on the plastic strain rate $\dot \varepsilon_\mathrm{p}$ which is computed according to
\begin{equation}
    \dot \varepsilon_\mathrm{p}(t_n) = \frac{\varepsilon_\mathrm{p}(t_n)-\varepsilon_\mathrm{p}(t_{n-1})}{t_n-t_{n-1}}
\end{equation}
 where $\varepsilon_\mathrm{p}$ is the accumulated plastic strain and $t_n$ is the elapsed time after $n$ timesteps. An avalanche is detected if the plastic strain rate exceeds a certain threshold $\dot{\varepsilon}_\mathrm{p,th}$. The box size naturally affects the extent of mobile dislocations as well as the noise level, therefore, different thresholds have to be applied for different box sizes. The thresholds are not dependent, however, on the extent of quenched disorder. The values of $\dot{\varepsilon}_\mathrm{p,th}$ are summarized in the supplementary material. The thresholds are chosen such that they are above the typical noise level but low enough to detect signals of significant size. The LYS assigned to a box is defined as the external stress at the onset of the first avalanche detected.

\section{Results}
\label{sec:results}
In order to investigate the effect of precipitates on the plastic behavior, two fundamentally different cases are considered. We study systems consisting of only dislocation lines and ones that (besides dislocations) include strong quenched disorder in the form of a high density of strong (impenetrable) precipitates. In the former case dynamics is purely determined by the elastic long-range interaction of dislocations and the external stress. The latter case, however, approximates well the other limiting scenario when the short-range dislocation-precipitate interactions dominate the dynamics. In the following the two scenarios are compared in terms of the plastic response and the LYSs.

\subsection{The effect of precipitates and box division on the plastic response}

A total of $20$ initial dislocation configurations are created for both types of systems with and without precipitates. The division and external loading is performed for all $20+20$ systems. Sample configurations stressed to $100\,\mathrm{MPa}$ are visualized in Fig.\,\ref{fig:congfigs2} with subfigure (a) representing the precipitate-free system and the volume subdivisions whereas subfigures (b)-(d) illustrate the impact of precipitates on the system. It should be noted that a small number of boxes, especially for $B=2$, some of the boxes contain insufficient or zero amount of dislocation segments, that lead to diverging LYSs. Thus, instead of the average, the median is chosen to obtain representative stress-strain curves.

\begin{figure}[!h]
    \centering
    \includegraphics[width=0.9\columnwidth]{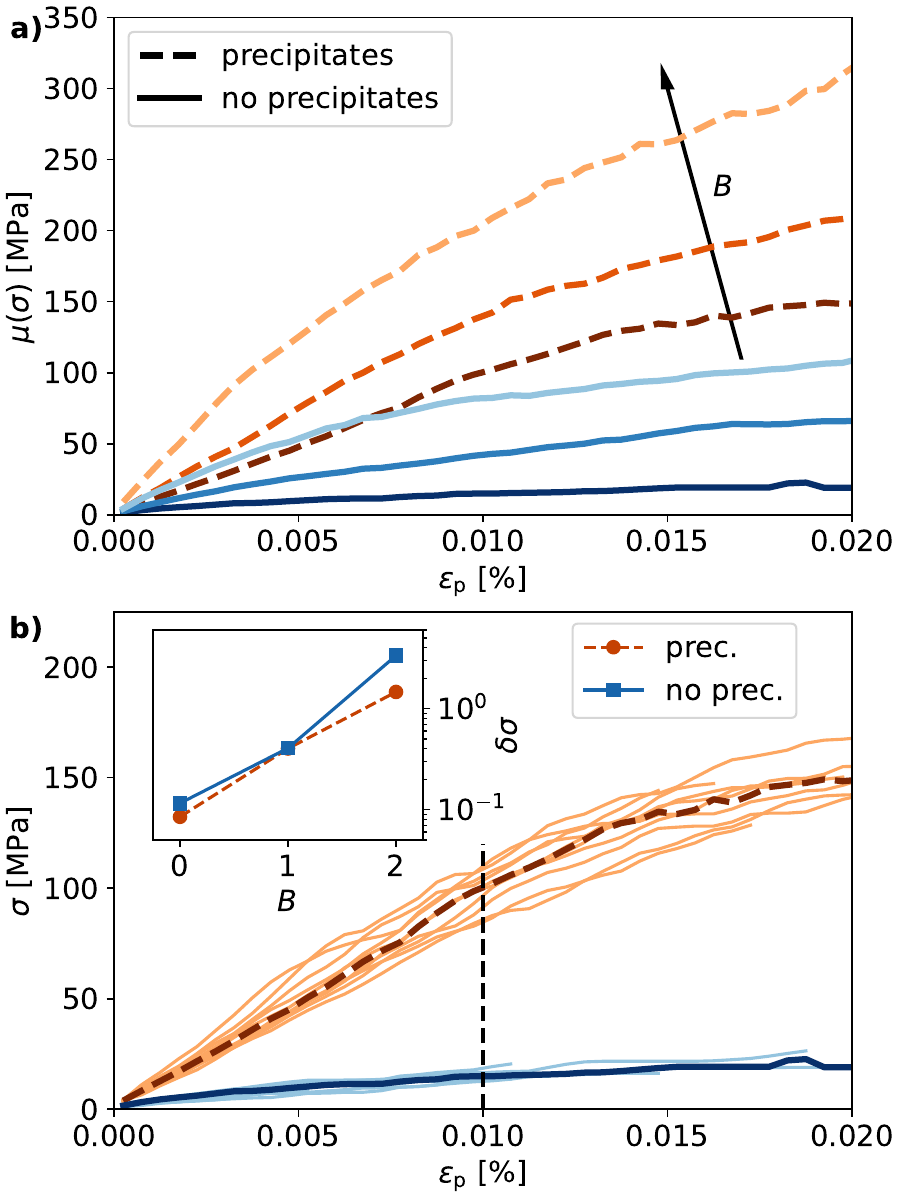}
    \caption{(a): the median stress $\mu(\sigma)$ as a function of the plastic strain $\varepsilon_\mathrm{p}$ computed for the ensembles of boxes of different sizes $B$. A clear hardening is observable if the box size is decreased (i.e. $B$ is increased) or precipitates are introduced to the system. (b): The individual stress-strain curves (pale-shade lines) exhibit visible fluctuations around the median behavior even for the $B=0$ cases plotted here. The inset shows the relative fluctuation $\delta \sigma$ normalized with the median value at $\varepsilon_\mathrm{p}=0.01\%$ (indicated by the black dashed line). The relative fluctuations are significantly higher in smaller boxes but are similar with or without precipitates.}
    \label{fig:stress-strain}
\end{figure}

The median stress-plastic strain curves show that the inclusion of precipitates significantly increases the strength of both the whole systems ($B=0$) and the subboxes ($B>0$) (see Fig.~\ref{fig:stress-strain}(a)). Additionally, a hardening size-effect can be observed as the box size is decreased (that is, $B$ is increased). Figure \ref{fig:stress-strain}(b) shows individual stress-strain curves for $B=0$ along the corresponding median curves. The relative fluctuation sizes, normalized with their median value, for $\varepsilon_\mathrm{p}=0.01\%$ are visualized in the inset. The relative fluctuations $\delta \sigma$ around the median curves increase as the box sizes are decreased, while the inclusion of precipitates has a little effect on the relative fluctuations. 

Figure \ref{fig:3d}(a) and (b) show the evolution of two plastic events at loading cases $B=0$ and $B=1$ for the same initial configuration consisting of dislocations and precipitates (which are not shown for visibility purposes). The two events clearly share very similar features in terms of dislocation activity. Furthermore, according to panel (c) these alike events occur at quite similar external stress values as well. This similarity implies that there might indeed be an underlying weakest link principle in the plastic behavior which motivates us to analyze whether and how the weakest link principle manifests and to compare the two scenarios, with and without precipitates.

\begin{figure}[!h]
    \centering
        \includegraphics[width=\columnwidth]{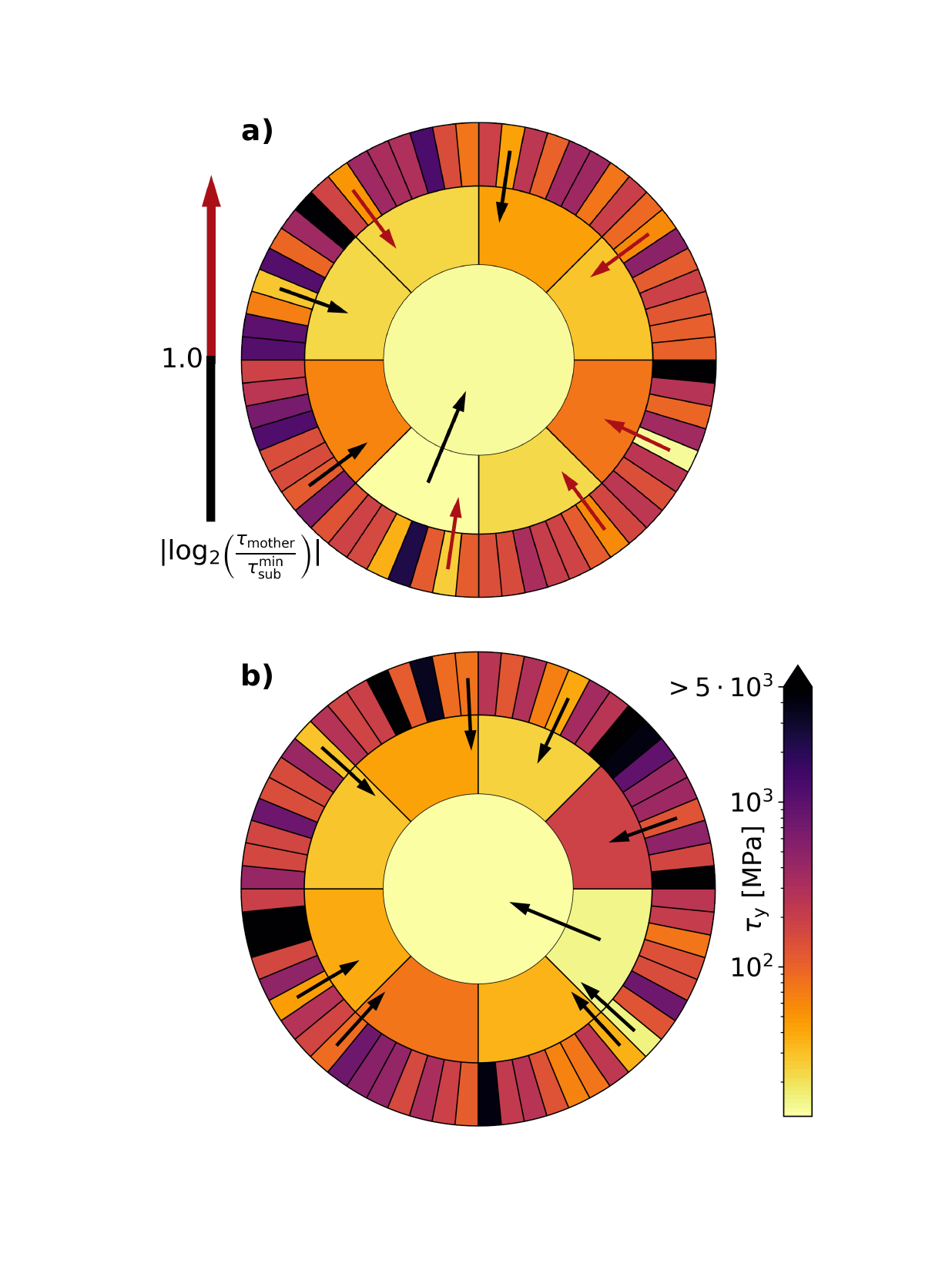}
    \caption{Tree charts representing the LYSs ($\tau_\mathrm{y}$) in two example systems. The three rings correspond to the three extents of division ($B=0$ is the inner circle and $B=2$ is the outer ring). The adjacency between cells of different rings indicates parent box - subbox relation. The inheritance of the LYSs from the softest subboxes is highlighted by the black arrows. Red arrows correspond to cases where the LYS of the softest subbox is not inherited to the parent box (there is at least a factor of 2 difference). (a): a configuration without precipitates in which the LYS inheritance is often violated. (b): a configuration with precipitates with particularly good inheritance behavior.}
    \label{fig:tree}
\end{figure}

\subsection{Local yield stress statistics}

The LYS values at different length-scales are shown for example configurations in Fig.~\ref{fig:tree}. These tree charts consist of an inner ring representing the full box $B=0$ surrounded by two outer rings subdivided into $8$ and $64$ cells for $B=1$ and $B=2$, respectively. Parent box - subbox relations are indicated by the adjacency of the cells. According to the weakest link principle, one expects the LYS of a box to be equal to the smallest LYS of its subboxes. Therefore, the arrows point from the subbox with the lowest LYS (softest subbox) to the parent box, while the arrow colors (black for good, and red for poor inheritance) quantify the relative difference in the yield stresses. Fig.~\ref{fig:tree}(a) represents a configuration (without precipitates) in which the expected inheritance of yield stresses is often violated. On the other hand, Fig.~\ref{fig:tree}(b) corresponds to an exemplary system (with precipitates) exhibiting perfect weakest link behavior. The tree charts of all the systems used for this study are shown in the supplementary material demonstrating that there is some variation how the weakest link principle manifests in individual configurations. Below we, therefore, continue with the statistical comparison of the two scenarios (with and without precipitates) in terms of LYSs and weakest link behavior.

\begin{figure}[!h]
    \centering
    \includegraphics[width=\columnwidth]{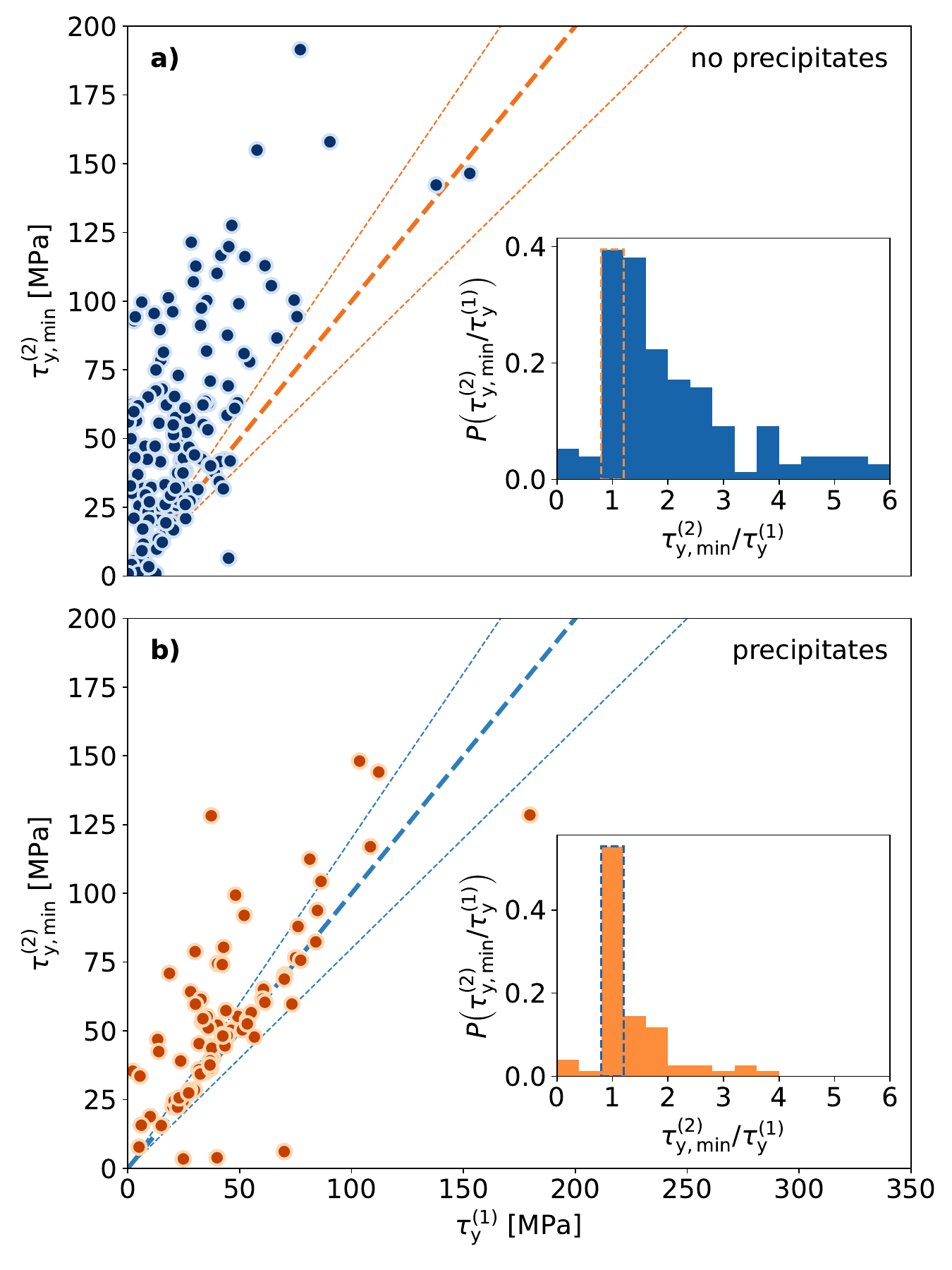}
    \caption{The LYSs $\tau_\mathrm{y,min}^{(2)}$ of the softest subboxes at scale $B=2$ against the LYSs $\tau_\mathrm{y}^{(1)}$ of the parent boxes at scale $B=1$ in systems (a) without precipitates and (b) with precipitates. The dashed lines correspond to the identity line with a $20\%$ tolerance marked by the dotted lines. The insets show the distribution of the ratio $\tau_\mathrm{y,min}^{(2)}/\tau_\mathrm{y}^{(1)}$ with a bin size of $0.4$ and the contoured bars corresponding to the points bounded by the dotted lines of the main figures.}
    \label{fig:scatter}
\end{figure}

\begin{figure}[!h]
    \centering
    \includegraphics[width=\columnwidth]{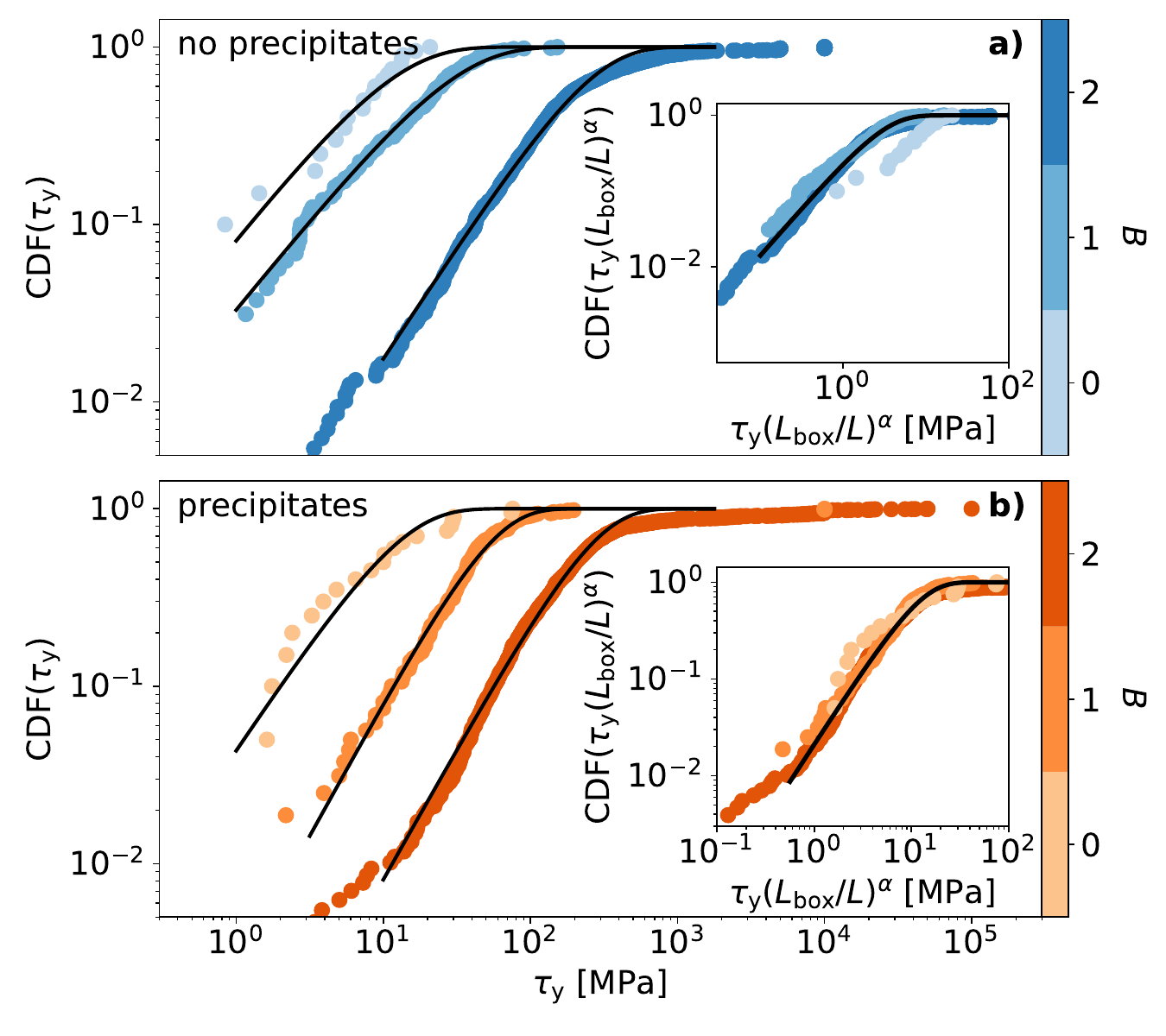}
    \caption{The cumulative distribution function (CDF) of the LYS $\tau_\mathrm{y}$ and the fitted Weibull distribution at different box sizes $B$  in systems (a) without precipitates and (b) with precipitates. Insets: Curve-collapse can be obtained by rescaling yield stresses with the box size $L_\mathrm{box}$ with exponent $\alpha$. For the values of $\alpha$ and the shape-parameters $k$ of the fitted principle curves see Tab.~\ref{tab:dimension}.}
    \label{fig:Weibull}
\end{figure}

Let us assume that the behavior of the system is determined by the weakest link principle and the strength of the links follows a distribution with a power law asymptotic behavior. More precisely, it is assumed that the probability density function $p$ of the yield stress $\tau_\mathrm{link}$ associated with a link starts as $p(\tau_\mathrm{link})\propto \tau_\mathrm{link}^k$ at $\tau_\mathrm{link}\rightarrow 0$. In this case the LYS $\tau_\mathrm{y}$ (at a certain scale $B$) is expected to obey Weibull distribution \cite{weibull1939statistical, weibull1951wide}:
\begin{equation}
    \mathrm{CDF}(\tau_\mathrm{y})=\mathrm{CDF}_\mathrm{Weibull}(\tau_\mathrm{y}, k, \lambda) = 1-\mathrm{exp}\left[\left(\frac{\tau_\mathrm{y}}{\lambda}\right)^k\right].
\end{equation}
Here, $k$ and $\lambda$ are the so-called \emph{shape parameter} and \emph{scale parameter}, respectively. The LYS data can indeed be fitted well with a Weibull distribution for all three scales considered, with or without precipitates (see Fig.~\ref{fig:Weibull}). In order to obtain a proper estimate of parameter $k$ the low-stress tail of the distribution should be fitted. Therefore, when fitting by employing the least-squares method each residual (corresponding to a data point) is weighted with a weight proportional to $1/\mathrm{CDF}(\tau_\mathrm{y})$. Curve collapse can be obtained by rescaling the LYS values with power $\alpha$ of the linear box size $L_\mathrm{box}$  (see the insets of Fig.~\ref{fig:Weibull}) and it can be fitted with a master curve.

It was shown earlier that the number of links is scaling with the linear size $L_\mathrm{box}$ with an exponent $D_\mathrm{link}=k\alpha$ (which we will refer to as link dimension) where $k$ is the shape-parameter of the master curve \cite{ispanovity2017role}. The link dimension values and the corresponding fit parameters are listed in Tab. \ref{tab:dimension}. In the case of configurations with precipitates $D_\mathrm{link}$ is very close to the dimension $D=3$ while $D_\mathrm{link}$ is significantly higher than $D$ if precipitates are not present. This can be interpreted as follows: if precipitates are included, links are more localized, therefore, an almost extensive scaling of links can be observed ($D_\mathrm{link}\approx D$). In the systems without precipitates, however, links can be highly extended. These links might not be activated in smaller boxes, therefore, as the box size is increased, virtually new activable links appear resulting in a super-extensive scaling ($D_\mathrm{link}>D$). From this result it can be concluded that when short-range interaction have a significant impact on the dislocation dynamics the traditional weakest link behavior can be observed. However, if the system is dominated by the long-range elastic interaction of dislocations, the dynamical behavior does not obey the conventional weakest link principle and the dynamics is more complex. 

The same can be concluded based on the statistical analysis of the parent box - subbox relations. To this end, we consider the softest subboxes for $B=2$, and compare their LYSs $\tau_\mathrm{y,min}^{(2)}$ with the values of the corresponding parent boxes, as shown in Fig.~\ref{fig:scatter}. The difference between the systems without precipitates [Fig.~\ref{fig:scatter}(a)] and with precipitates [Fig.~\ref{fig:scatter}(b)] is remarkable. The distribution of the ratio of the LYSs of the softest subboxes and their parent boxes is shown in the insets. In systems without precipitates about $20\%$ of the parent boxes inherit the yield threshold from their softest subbox within a precision of $20\%$. In the case of configurations containing precipitates, however, this precise inheritance is more prevalent: it is true for more than half of the parent boxes. In an ideal weakest link scenario (in which the LYSs are accurately inherited from lower scales) the Spearman correlation $C_{1,2}^\mathrm{s}$ between the local yield thresholds of parent boxes at scale $B=1$ and their softest subboxes at scale $B=2$ would be exactly 1. The ratio of yield stresses of the softest subbox and the parent box would also be exactly 1. Let us denote the median of this ratio with $\mu_{1,2}$ which, obviously, should be also 1 in an idealistic case. The results indicate that the configurations with precipitates behave more similarly to the ideal weakest link scenario in terms of $C_{1,2}^\mathrm{s}$ and $\mu_{1,2}$ as well (see Tab.~\ref{tab:weakestlink}). 

\begin{table}[H]
\centering
\begin{tabular}{lccc}
\hline\hline
 & $k$ & $\alpha$ & $D_\mathrm{link}$ \\ \hline
no precipitates & $1.25\pm0.05$ & $3.20\pm0.10$ & $4.0\pm0.3$ \\
precipitates & $1.55\pm0.05$ & $2.20\pm0.05$ & $3.4\pm0.2$ \\ \hline\hline
\end{tabular}
\caption{The shape-parameters $k$ of the principle curves, the exponents $\alpha$ used for obtaining curve-collapse (see Fig.~\ref{fig:Weibull}) and the derived link dimensions $D_\mathrm{link}=k\alpha$. Systems with and without precipitates exhibit extensive and super-extensive scaling of link numbers, respectively.}
\label{tab:dimension}
\end{table}

\begin{table}[H]
\centering
\begin{tabular}{lcc}
\hline\hline
 & $C_{1,2}^\mathrm{s}$ & $\mu_{1,2}$ \\ \hline
no precipitates & $0.42$ & $2.00$  \\
precipitates & $0.72$ & $1.12$  \\
weakest link & $1.00$ & $1.00$  \\ \hline\hline
\end{tabular}
\caption{Statistical measures characterizing the relationship between the LYSs of parent boxes at scale $B=1$ and their softest subboxes at scale $B=2$ for our simulations (with and without precipitates) and for an idealistic weakest link scenario. $C_{1,2}^\mathrm{s}$ and $\mu_{1,2}$ are their Spearman correlation and the reciprocal of the median of their ratio, respectively.  }
\label{tab:weakestlink}
\end{table}

\begin{figure}[!h]
    \centering
    \includegraphics[width=\columnwidth]{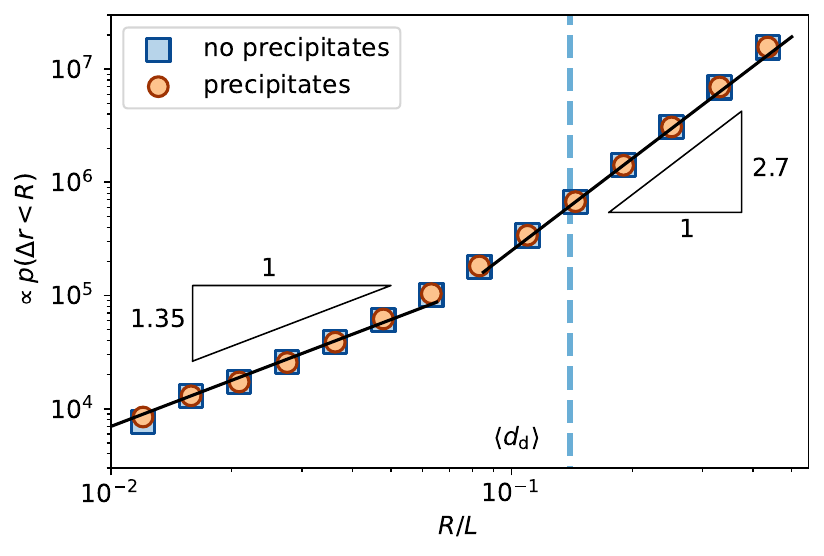}
    \caption{The correlation integral of dislocation nodes, that is, the probability of two nodes being closer than a certain distance $R$. $\Delta r$ and $L$ are the distance of two nodes and the linear system size, respectively. The configurations with and without quenched disorder yield the same correlation integral with the same characteristic static length scale around the half of the mean dislocation spacing $\langle d_\mathrm{d}\rangle$ indicated by the vertical blue dashed line. Below that the fractal dimension of the dislocation structure is close to one suggesting an essentially 1D (line-like) anatomy. Above the static length scale the fractal dimension is close to the system dimension $D=3$ indicating that the structure is more complex on larger scale but still somewhat below $D$ due to the spatial correlation of dislocations.}
    \label{fig:static_length_scale}
\end{figure}

\begin{figure}[!h]
    \centering
    \includegraphics[width=\columnwidth]{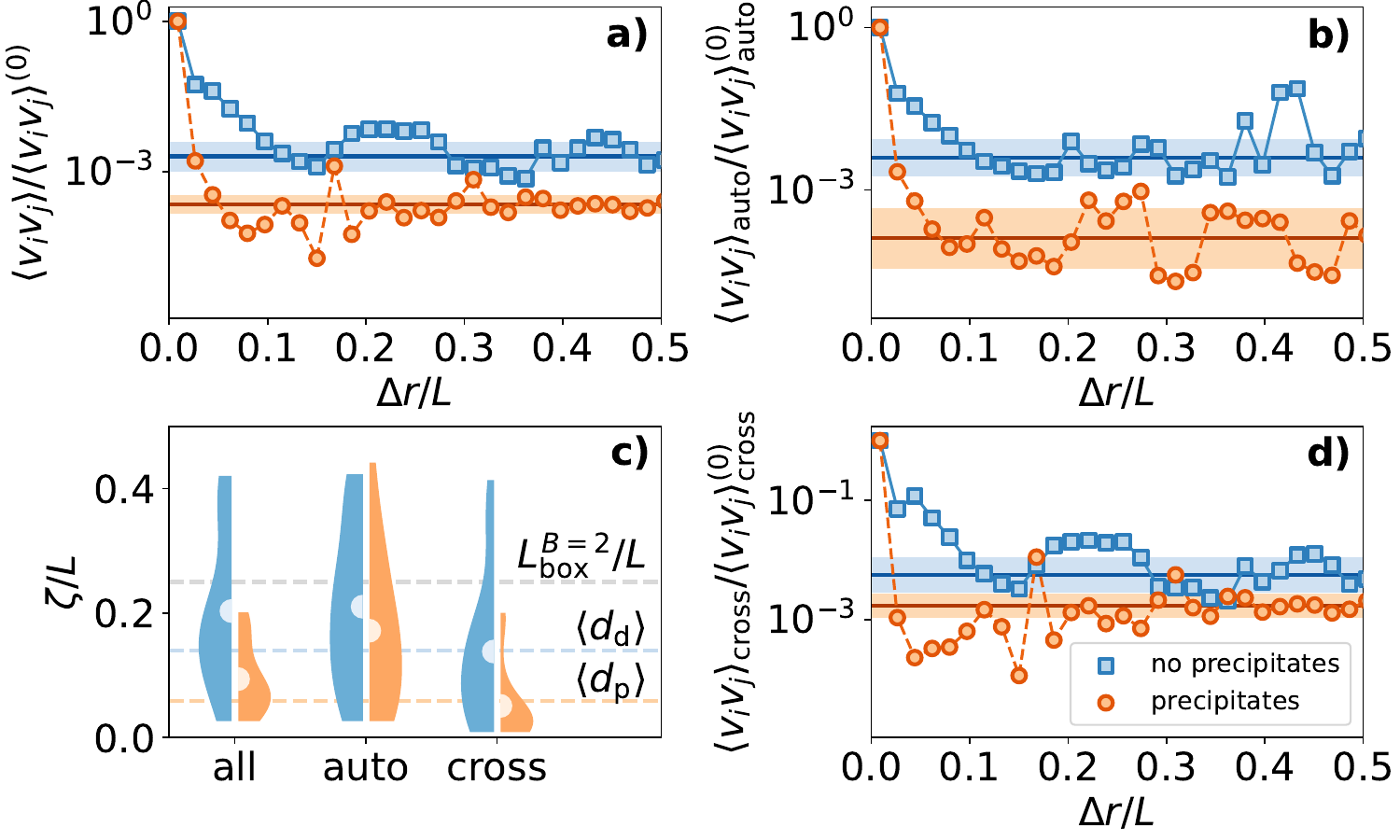}
    \caption{(a): The nodal velocity correlation $\langle v_iv_j\rangle$ at the first avalanches at their onset averaged for the two ensembles of configurations (with and without precipitates). The data is binned w.r.t. the node distance $\Delta r$ (normalized with the linear size $L$ of the simulation cell) and is normalized with the value $\langle v_iv_j\rangle^{(0)}$ of the first bin. The horizontal lines and stripes denote the noise level to which the velocity correlation tends for pairs of remote dislocation nodes and  its extent of fluctuations, respectively. The correlations for node pairs of the same dislocation lines and of distinct dislocation lines are shown in panel (b) and (d), respectively. (c): the distribution of the length-scale $\zeta$ at which $\langle v_iv_j\rangle$ blends into the noise in individual configurations. The mode of the distribution (for all nodes) coincides well with the mean dislocation spacing $\langle d_\mathrm{d}\rangle$ and with the mean precipitate spacing $\langle d_\mathrm{p}\rangle$ (colored dashed lines) in the no precipitates and precipitates cases, respectively. The light half circles denote the mean values of $\zeta/L$ and the gray dashed line indicates the linear size of the smallest ($B=2$) boxes employed for LYS analysis.}
    \label{fig:triggering_length_scale}
\end{figure}

\subsection{Length scales}

So far the LYS statistics was analysed in terms of its connection to the weakest link principle. Now we turn to length scales, another set of important ingredients of the mesoscopic modeling of plasticity, and relate them to how the weakest link principle manifests in systems with and without precipitates. In order to characterize the static correlations of the dislocation configurations the correlation integral of dislocation discretization nodes is computed for the two ensembles of configurations. The value of the correlation integral at distance $R$ is proportional to the probability of a pair of nodes being closer than $R$. As it is shown in Fig.~\ref{fig:static_length_scale} the correlation function signifies the existence of two regimes both exhibiting a scaling with $R$. For low $R$ values the scaling exponent is close to 1 suggesting a practically 1D (line-like) structure. For large $R$ values the scaling exponent approaches the system dimension $D=3$ signifying a more complex structure but it is still clearly below $D$ due to spatial correlation of dislocations. The data suggest that not only the length scale corresponding to the crossover between the two regimes is the same for the precipitate and no precipitate cases but even the whole $R$-dependent correlation integral is practically identical. That is, the presence of precipitates does not seem to influence the systems in terms of static correlations.

Despite the static correlations being the same, the different LYS statistics and weakest link behavior implies that precipitates significantly affect the dynamics. This will be demonstrated below using the analysis of dynamic correlation length $\zeta$. To this end, the nodal velocity correlation (for a single configuration) is defined as 
\begin{equation}
   \langle v_iv_j\rangle( \Delta \bm r) =  \left\langle\sum_{i=1}^{N_n}\sum_{\substack{j=1, \\ j\neq i}}^{N_n}{\delta(\Delta \bm r - \bm r_i + \bm r_j)v_iv_j}  \right\rangle
\end{equation}
where $\bm r_i$ and $v_i$ are the position and the magnitude of the velocity of the $i$th node, respectively, and $N_n$ is the total number of nodes. $\delta$ denotes the delta function. The ensemble averaged velocity correlation as a function of $\Delta r=|\Delta \bm r|$ is shown in Fig.~\ref{fig:triggering_length_scale}(a). The high values for small distances $\Delta r$ is consistent with the intuition that quickly moving nodes are surrounded by nodes that typically also have high velocities. As the distance $\Delta r$ is increased, the velocity correlation $\langle v_i v_j\rangle$ decreases until it blends into the noise level. We associate the dynamic length scale $\zeta$ with the smallest distance $\Delta r$ at which $\langle v_i v_j\rangle$ drops below the noise level. In order to distinguish the contribution of the movement of single dislocation lines (and consequently, the coordinated movement of the nodes of these lines) and the coherent motion of several lines, $\langle v_i v_j\rangle$ is decomposed into two terms:
\begin{equation}
    \langle v_i v_j\rangle = \langle v_i v_j\rangle_\mathrm{auto} + \langle v_i v_j\rangle_\mathrm{cross}
\end{equation}
where $\langle v_i v_j\rangle_\mathrm{auto}$ and $\langle v_i v_j\rangle_\mathrm{cross}$ are the velocity correlations of the pairs of nodes from the same and from different dislocation lines, respectively. Here, we consider two nodes to be of the same dislocation line if one node can be reached from another through connected dislocation segments while avoiding dislocation junctions. The average distance dependence of $\langle v_i v_j\rangle_\mathrm{auto}$ and $\langle v_i v_j\rangle_\mathrm{cross}$ are presented in Fig.~\ref{fig:triggering_length_scale}(b) and (d). The distribution of the dynamic length scale $\zeta$ corresponding to all the node pairs and the two subsets of the pairs is summarized in Fig.~\ref{fig:triggering_length_scale}(c). There is a clear difference in the velocity correlations between systems with and without precipitates in particular in the cross-correlation term: the dynamic length scale is larger on average and reaches larger values in the case of pure systems without precipitates. In terms of the total $\langle v_i v_j\rangle$ this translates to similar differences as in the case of cross-correlations mainly due to the fact that cross-correlation has much larger (on average $\sim50$ times larger) weight in $\langle v_i v_j\rangle$ than the autocorrelation.

\section{Discussion and conclusions}
\label{sec:conclusions}

To elucidate the relevance of our results in scale bridging between discrete and continuum dislocation descriptions of plasticity, let us first step back and look at the problem from a wider perspective. Material microstructures are often inhomogeneous and characterized by some kind of internal disorder. The origin of this disorder may have various forms, such as density fluctuations (like in fluids or amorphous materials) or a random distribution of lattice defects in crystalline materials (like variations in the grain structure or dislocation patterns). Intuitively one may think that microstructural disorder degrades mechanical properties and it should be avoided/reduced. However, the truth is often the opposite: as it is well-known from the Hall-Petch and Taylor relationships, increased amount of grain boundaries (i.e., smaller grains) and/or dislocations increase the strength of the material. But not only the amount of defects matters, but also their spatial fluctuations. For instance, it has been shown in amorphous systems that the increase of the fluctuations of the local atomic density can significantly increase ductility \cite{tuzes2017disorder, popovic2018elastoplastic}. Similarly, fluctuations in the lattice distortions due to varying atomic radii in random alloys (especially in high entropy alloys) increase strength by obstructing dislocation motion \cite{varvenne2016theory}. It has been shown by Rodney and co-workers that in order to understand the yielding phenomena in random alloys, one needs to understand the statistical properties (distribution and spatial correlations) of the microstructural disorder \cite{geslin2021microelasticity, geslin2021microelasticity2, rodney2024does}.

In the case of dislocation-mediated plasticity, contrary to the two examples above, disorder does not originate from the atomic structure, but rather from the randomness of the configurations of the dislocation lines. For example, regions with higher dislocation density are typically stronger than those with lower dislocation content. Our main assumption motivating the presented research is that, similarly to other disordered materials, understanding mechanical properties necessitates the in-depth understanding of the statistical properties of the disorder. The next question we ask is how to quantify disorder for a complex dislocation microstructure. As it was elucidated in the introduction, dislocation avalanches are the direct manifestation of the microstructural disorder through the local variation of local strength. In this paper we, therefore, analyze the statistical properties of the local strength, and how it affects the nature of plastic deformation of crystals, especially when it comes to the relevant dynamical length scales. Here the key idea is to study to what extent the weakest link picture (that is, the inheritance of LYSs from lower to higher scales) is realized in dislocation systems with and without precipitates. We find that the sub-volumes triggered during the onset of the first avalanche tend to get confined by quenched disorder, and hence the crystal can be partitioned into sub-volumes within which a LYS value can be defined such that the weakest link picture works remarkably well in crystals with precipitates. On the other hand, the onset of avalanches in pure dislocation systems without any quenched pinning tends to be significantly less localized and typically involves several dislocations in different parts of the system. This then implies that, at least for the system sizes we are able to consider here, defining LYSs for scales smaller than the system size does not result in a clear-cut weakest link type of behavior.

It is important to note that the length scale characterizing the size of the triggered sub-volumes does not coincide with the static length scale of the system. In fact, the static correlations of the snapshots of configurations is practically the same for the two ensembles of configurations (with and without precipitates) studied in this work and it is the dynamic length scale (based on dislocation velocity correlations) what characterizes the extension of avalanche triggerings.

When it comes to scale bridging between discrete and continuum descriptions of plasticity it is important to note that fluctuations related to dislocation avalanches are usually averaged out in CDD descriptions, since yield strength is assumed to be a deterministic function of the local dislocation density via the Taylor relationship \cite{hochrainer2014continuum, xia2015preliminary, sudmanns2019dislocation, zoller2021microstructure}. As such, CDD models by construction cannot account for dislocation avalanches and the effect of small-scale microstructural disorder. To fill the gap and to understand the role of disorder in yielding, mesoscale stochastic continuum dislocation dynamics (SCDD) models have been introduced, where disorder is represented in the LYS that is assumed to be a random variable \cite{zaiser2005fluctuation, ispanovity2020emergence, wu2021cell}. To parameterize such models two fundamental questions need to be answered: how to choose (i) the spatial resolution of the model and (ii) the distribution of the LYS. The results presented here aim to address both these issues.

According to our results, when choosing the correct resolution for the mesoscopic modeling of plasticity, the dynamic length scale has to be considered instead of the static length scales (e.g., the typical dislocation spacing or characteristic distances related to the cell structure, etc.); otherwise the avalanche activity might not be modeled properly. In other words, this dynamic length scale determines the size of the representative volume element of the mesoscopic modeling of avalanche phenomena in the system.

Besides a properly chosen length scale, the LYS is also crucial ingredient of a transition to mesoscopic modeling as the heterogeneity originating from disorder is introduced by the variation of the LYS. Therefore, it is important to understand how exactly these stresses should be introduced in terms of, e.g., distribution, spatial correlations, etc. Our results show that the LYS is Weibull-distributed (as expected based on weakest link principles). While the present data is insufficient to acquire meaningful conclusions about spatial correlations, it can be assumed that at least in short range (i.e., for highly overlapping sub-volumes) the LYSs are not uncorrelated. Since the dynamic length scale characterizes the size of easily triggerable sub-volumes (soft spots), it is a natural assumption that it is connected to how the LYSs are correlated in space. However, it remains to be answered by future research whether and how the correlation length of LYSs is related to the dynamic length scale.

It should be emphasized that the present analysis is based on the ``first avalanches'' which occur at rather small stresses/strains, that is, typically well below the critical stress $\sigma_\mathrm{c}$ of the depinning transition of the dislocation assembly in the presence of strong enough precipitates~\cite{salmenjoki2020plastic}. Moreover, one has to also make a distinction between the length scales associated with the initial, triggering part of the avalanche, and that of the avalanche as a whole~\cite{berta2024avalanche}. This becomes evident especially when approaching $\sigma_c$ from below: The cutoff avalanche size (when considering an ``extensive'' measure of the avalanche size~\cite{ispanovity2014avalanches}) and hence the dynamic length scale defined by the spatial extent of the whole avalanche is expected to increase, and eventually diverge at $\sigma=\sigma_\mathrm{c}$ (this is also suggested by the results in Ref.~\cite{berta2024avalanche}). Thus, it would be interesting to study the onset of the large, system-spanning avalanches at the depinning critical point in 3D DDD systems with precipitates, to check if such avalanches are still triggered locally even if the avalanche as a whole would span the system. Such analysis would then give insight on the question of if the weakest link picture would break down even in the system with precipitates at the depinning critical point.

\section*{CRediT authorship contribution statement}
\noindent
{\bf D\'{e}nes Berta:} Conceptualization, Data curation, Investigation, Methodology, Visualization, Writing - original draft.
{\bf David Kurunczi-Papp:} Conceptualization, Data curation, Investigation, Methodology, Writing - original draft.
{\bf Lasse Laurson:} Conceptualization, Supervision, Writing - original draft.
{\bf P\'{e}ter Dus\'an Isp\'{a}novity:} Conceptualization, Supervision, Writing - original draft.

\section*{Declaration of competing interests}

\noindent
The authors declare that they have no known competing financial interests or personal relationships that could have appeared to influence the work reported in this paper.

\section*{Acknowledgements}

\noindent
DB and PDI acknowledge the financial support from the National Research, Development and Innovation Fund of Hungary under the young researchers’ excellence programme NKFIH-FK-138975. The authors wish to acknowledge CSC – IT Center for Science, Finland, for computational resources. DKP and LL acknowledge the financial support of the Research Council of Finland via the Academy Project COPLAST (Project No.~322405).

\section*{Supplementary materials}
\noindent
Supplementary material associated with this article can be found, in the online version, at [url will be inserted by the publisher].

\end{document}